\documentclass[prb,aps,twocolumn,superscriptaddress]{revtex4-2}

\usepackage{graphicx}
\usepackage{dcolumn}
\usepackage{bm}
\usepackage{comment}
\usepackage{xcolor}
\usepackage{amsmath}
\usepackage{amssymb}

\begin{document}

\preprint{APS/123-QED}


\title{Quantum geometric tensor in systems with fractional band dispersion}

\author{Jamme Omar A.~Biscocho}
\email{jabiscocho@up.edu.ph}
\author{Kristian Hauser A.~Villegas}%
 \email{kavillegas1@up.edu.ph}
\affiliation{%
 National Institute of Physics, University of the Philippines Diliman, Philippines.\\
}%

\date{\today}

\begin{abstract}
We investigate the quantum geometric tensor, which is comprised of the Berry curvature and quantum metric, in a generalized Dirac two-band system with non-integer dispersion $E(\mathbf{k})\sim k^{\alpha}$. Our analysis reveals that this type of dispersion introduces significant and novel effects on quantum band geometry. We calculate the Berry curvature and observe its redistribution in momentum space as \(\alpha\) varies. Notably, despite this redistribution, the change in Chern number across topological transitions remains quantized as an integer, even for non-integer \(\alpha\). We illustrate the physical implications of this redistribution by computing the orbital magnetization. Furthermore, we demonstrate that the Berry curvature and quantum metric peak along the regions of momentum space where the energy band exhibits high curvature. While it is well-established that Berry curvature becomes highly concentrated at the band touching points, our findings indicate that they can also accumulate at sharp band corners, away from these band touching points. We further show that while the Berry curvature develops Dirac delta-like peaks away from $\mathbf{k}=0$, the Berry monopole, corresponding to the topological charge, remains located at the origin.
\end{abstract}

\maketitle

\section{Introduction}\label{sec:intro}
The energy-momentum dispersion with integer exponent $E(\mathbf{k})\sim k^n$ is ubiquitous in physics. Notable examples include the nonrelativistic free particle with 
$n=2$, Dirac dispersion with 
$n=1$, and flat bands with 
$n=0$. In solid-state systems, where the dispersion relations are often trigonometric functions of the crystal momentum 
$\mathbf{k}$, integer-exponent dispersions arise when the energy is expanded around the band minimum. This is exemplified by the linear dispersion in graphene and the quadratic dispersion observed in semiconductors and metals. Moreover, it has been demonstrated that an N-layer ABC-stacked graphene exhibits a dispersion relation of the form 
$E(\mathbf{k})\sim k^N$ \cite{Zhang2010}.

A more exotic category of dispersion relations involves systems with non-integer exponents $\alpha$ in the form $E(\mathbf{k})\sim k^\alpha$. This behavior is observed in a variety of contexts, including networks with long-range transport \cite{RiascosNetworks2015}, nonlocal field theories \cite{AtmanNonlocFields2022}, optical phenomena in inhomogeneous linear media \cite{IominGraviOptics2021, Zakeri2023}, and fluctuations at the domain wall in superfluid-superfluid interfaces, which exhibit a dispersion relation of $E(\mathbf{k})\sim k^{3/2}$ \cite{Watanabe2014, Watanabe2020}. In bilayer graphene systems, the fractional dispersion can interpolate between $\alpha=2$ and $\alpha=4$, with the exponent $\alpha$ being tunable by the displacement field applied perpendicular to the layers \cite{Dong2023}. Modifying the non-integer exponent has been shown to have significant effects on various phase transitions in both bilayer and trilayer graphene systems \cite{Raines2024}.

Fractional dispersion can also be interpreted as the momentum space representation of fractional quantum mechanics \cite{LaskinFQM, HerrmannFC}. This framework has found diverse applications, including in entropy-area relations in black hole thermodynamics \cite{JalalzadehBHThermo2021} and in the confinement of quarks \cite{HerrmannFC, Abu-ShadyFractQuark2024}.

In real-space lattices, space-fractional quantum mechanics can be realized by extending the tight-binding model to incorporate infinite-range hopping, then using the Gr\"unwald-Letnikov form of the fractional derivative \cite{Dartora2021}. This approach allows the resulting tight-binding Hamiltonian to define a L\'evy crystal \cite{Stickler2013}. Such fractional media have been shown to support a variety of novel phenomena distinct from conventional quantum mechanics, including localization-delocalization transitions in one-dimensional systems \cite{Chatterjee2023}, exotic soliton-like wave packet evolution in optical experiments \cite{Zakeri2023}, and the breakdown of the Mermin-Wagner-Hohenberg theorem \cite{Dartora2024}. Additionally, the effects of Coulomb interactions have been explored in fractional Dirac materials, demonstrating that the fractional dispersion is robust due to its non-analytic structure \cite{Roy2023}.

In recent years, the focus of band theory in condensed matter physics has shifted from topology to geometry, effectively encapsulated by the quantum geometric tensor. The real part of this tensor, known as the quantum metric \cite{Provost1980}, measures amplitude distance, while its imaginary part, the Berry curvature, quantifies changes in phase. The Berry curvature has been extensively studied and serves as a cornerstone of modern theories of states of matter \cite{Resta1993, Vanderbilt1993, Ceresoli2006, Resta2011}. There has also been a growing interest in understanding the nature and physical implications of the quantum metric \cite{Torma2024}. Early investigations primarily focused on superfluid weight in the superconducting phase of twisted bilayer graphene \cite{Hu2019, Tian2023, Julku2020}. More recent studies have highlighted the crucial role of geometry in the superconducting pairing potential \cite{Kitamura2023, Kitamura2024, Daido2024}, as well as its effects on collective modes in superconductors \cite{Villegas2021, Villegas2023} and the enhancement of electron-phonon coupling \cite{Yu2024}.

Most existing research has shown how quantum geometry influences various phenomena, including transport, superconductivity, and optical responses, primarily using conventional integer-exponent band dispersion. However, the emergence of non-integer energy band dispersions across different platforms highlights the need for a comprehensive investigation of their band geometry. In this work, we demonstrate that non-integer dispersion results in novel and significant effects on quantum band geometry.

This manuscript is organized as follows. In Sec. \ref{sec:theory}, we introduce our model Hamiltonian and discuss the behavior and oddities of its energy bands. In Sec. \ref{sec:berry}, we investigate the Berry curvature and the change in Chern number across a possible topological transition. To explore the physical consequence of the Berry curvature redistribution, we calculate the orbital magnetization in Sec. \ref{sec:orbital}. We then proceed to the investigation of the quantum metric in Sec. \ref{sec:metric}. Lastly, we give our conclusions in Sec. \ref{sec:conclusion}.

\section{Hamiltonian Model}\label{sec:theory}
\subsection{Fractional dispersion from tight-binding}
In this section, we review the emergence of fractional dispersion from a one-dimensional tight-binding model, called L\'evy crystals \cite{Stickler2013}, following the discussion in \cite{Dartora2021}. We then outline how this can be extended to higher dimensions and to topological insulators.

Our starting point is the Gr\"unwald-Letnikov definition of the fractional derivative
\begin{align}
\label{glderivative}
     \frac{d^\alpha f(x)}{dx^\alpha}=\lim_{\epsilon\rightarrow 0}\frac{1}{\epsilon^\alpha}\sum_{n=0}^\infty\frac{(-1)^k\Gamma(\alpha+1)}{\Gamma(\alpha-n+1)n!}f(x-n\epsilon).
 \end{align}

Now, consider a tight-binding Hamiltonian
 \begin{align}
 \label{htight}
     H=\sum_{ij}t_{ij}c^\dagger_ic_j,
 \end{align}
 with the hopping given by
 \begin{align}
 \label{tfrac}
     t_{ij}=(-1)^{|i-j|}\frac{\Gamma(\alpha+1)}{\Gamma(\alpha-|i-j|)(|i-j|+1)!}t_0,
 \end{align}
where $i$ and $j$ label the lattice sites. If we substitute Eq. \eqref{tfrac} into the Hamiltonian \eqref{htight}, we get a summation of the form similar to Eq. \eqref{glderivative}. Taking the infinite lattice limit $N\rightarrow\infty$ and long wavelength limit $\epsilon\rightarrow0$, where $\epsilon$ is the lattice constant, precisely gives us the Gr\"unwald-Letnikov fractional derivative, which in the momentum space gives $E(k)\sim k^\alpha$. We note that \( \alpha \) can be any real number and is not limited to fractional values. Nevertheless, we adopt the conventional terminology and refer to \( \alpha \) as \lq\lq fractional exponent.\rq\rq

We can extend this construction to get a multiband system by adding the orbital indices $a$ and $b$, and write Eq. \eqref{tfrac} as
\begin{align}
 \label{tfracmultiband}
     t_{ia,jb}=(-1)^{|i-j|}\frac{\Gamma(\alpha+1)}{\Gamma(\alpha-|i-j|)(|i-j|+1)!}(t_0)_{ab},
 \end{align}
Note that the hopping scale $(t_0)_{ab}$ is now a matrix in orbital space, but the hopping across different lattice sites has the same profile as Eq. \eqref{tfrac}, thereby still giving us a fractional derivative in the long wavelength limit.

We can further extend this construction to higher dimensions by simply adding perpendicular directions. For instance, to obtain a two-dimensional L\'evy crystal, we introduce hopping along the \(y\)-direction in addition to the \(x\)-direction. The lattice index \(i\) is then generalized to a two-dimensional vector \(\mathbf{R}_i\). Furthermore, a L\'evy crystal topological insulator can potentially be realized by incorporating a phase factor \(\exp(i\phi_{ia,jb})\) in the hopping term in Eq. \eqref{tfracmultiband}, creating a flux through a given plaquette of the lattice, similar to the Haldane model \cite{Haldane1988}. However, in this work, we focus on the low-energy effective Hamiltonian with fractional dispersion and defer the explicit tight-binding construction to future studies.

\subsection{Fractional Dirac Hamiltonian}
We start with a simple model featuring non-integer dispersion and topological transition. The minimal model capable of achieving this is a two-band system, prompting us to extend the Dirac Hamiltonian into a Bloch Hamiltonian with non-integer dispersion:
\begin{align}
\label{hamiltonian}
H_\alpha(\mathbf{k}) = \text{sgn}(k_x)|k_x|^{\alpha}\sigma^x + \text{sgn}(k_y)|k_y|^{\alpha}\sigma^y + m\sigma^z.
\end{align}
In this expression, \(\mathbf{k}\) denotes the momentum, \(m\) represents the band gap parameter, and \(\sigma^x\), \(\sigma^y\), and \(\sigma^z\) are the Pauli matrices in the sublattice basis. The absolute values for the momentum components ensure that the Hamiltonian remains single-valued. This model generalizes the one-band case \cite{HerrmannFC} and is closely related to the framework presented in Ref. \cite{Roy2023}. Notably, when \(\alpha = 1\), the Hamiltonian reduces to the Dirac Hamiltonian, which undergoes a topological phase transition as the gap \(m\) varies from negative to positive. 

Additionally, we note that the fractional Dirac equation can be derived in real space through a generalized factorization of the Schrödinger equation, where \(\alpha = 2/N\) characterize an internal SU(N) structure of the particle \cite{HerrmannFC}.

\subsection{Band structure and Bloch functions}
\begin{figure*}[htb!]
	\centering
	\includegraphics[width=1\linewidth]{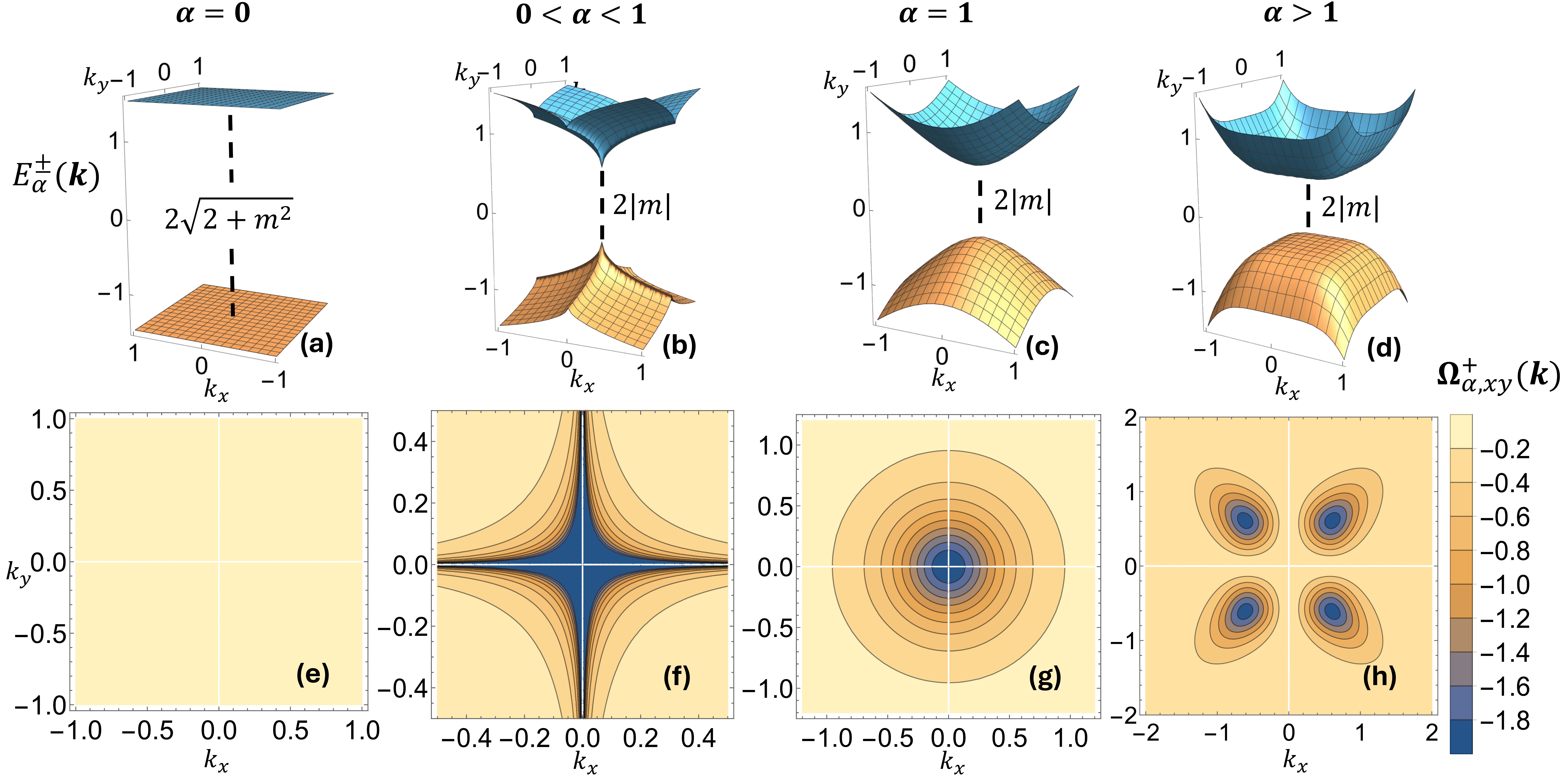}
	\caption{The plots depict the energy bands (a)-(d) and the Berry curvature (e)-(h) of the conduction band for various values of \(\alpha\) with \(m=0.5\). In panels (b) and (f), we set \(\alpha=0.3\), while in panels (d) and (h), \(\alpha=2.5\). The expressions next to the dashed lines in panels (a)-(d) indicate the magnitude of the band gap.}
	\label{fig1:results}
\end{figure*}
Our model in Eq. \eqref{hamiltonian} has the advantage that while it is sufficiently complex to exhibit nontrivial phases, it is also simple enough to allow analytical calculations of the Berry curvature and quantum metric. To start, the eigenstates are
\begin{align}
\label{eigenstates}
\psi^{_\pm}_{\alpha (\mathbf{k})}=&\frac{1}{\sqrt{2d(\mathbf{k},\alpha)[d(\mathbf{k},\alpha)\pm m]}}\nonumber\\&\times
\begin{pmatrix}
    m\pm d(\mathbf{k},\alpha) \\
    d_1(\mathbf{k},\alpha)+id_2(\mathbf{k},\alpha)
\end{pmatrix},
\end{align}
which correspond to the eigenvalues
\begin{align} \label{eq:eigenvalues}
E^\pm_{\alpha}(\mathbf{k})=&\pm\sqrt{d_1(\mathbf{k},\alpha)^2+d_2(\mathbf{k},\alpha)^2+m^2}\\
\equiv &\pm d(\mathbf{k},\alpha),
\end{align}
where 
\begin{align}
\label{d1}
   d_1(\mathbf{k},\alpha)&=\text{sgn}(k_x)|k_x|^{\alpha}\\ 
\label{d2}
   d_2(\mathbf{k},\alpha)&=\text{sgn}(k_y)|k_y|^{\alpha}.
\end{align}

In Fig. \ref{fig1:results} (a)-(d), we present the two bands for various values of \(\alpha\). For \(\alpha = 0\), we observe two flat bands, while \(\alpha = 1\) corresponds to a massive Dirac band (with \(m \neq 0\)). For all positive values of \(\alpha\), a zero value of \(m\) leads to a closing of the gap between the upper and lower bands. Specifically, for these values of \(\alpha\), the band gap is given by \(\Delta E = 2|m|\).

In contrast, the case of \(\alpha = 0\) behaves differently; there is no gap closing when \(m = 0\) because the band gap is given by \(\Delta E = 2\sqrt{2 + m^2}\). Later, when we analyze the change in Chern number, we will show that this case does not exhibit a topological phase transition, in contrast to the scenarios where \(\alpha > 0\).

We further observed that non-integer values of \(\alpha\) break the continuous rotational symmetry of the band structure in momentum space around \(\mathbf{k} = 0\). For \(0 < \alpha < 1\), the energy bands exhibit non-analytic behavior along the \(k_x\) and \(k_y\) axes as shown on Fig. \ref{fig1:results} (b). In contrast, for \(\alpha > 1\), shown in Fig. \ref{fig1:results} (d) the energy band surfaces take on a flat, square profile in the vicinity of the origin, \(\mathbf{k} = 0\). These energy band surfaces become more rectangular with four sharp corners for larger $\alpha$.

A consequence of the non-analyticity of the energy bands along the $k_x$ and $k_y$ axes for $0<\alpha<1$ is the divergence and discontinuity of the Fermi velocity,
\begin{align} \label{eq:velocity}
\mathbf{v}_{F}^{\pm}(\mathbf{k}, \alpha) = \mathbf{\nabla}_{\mathbf{k}}E^{\pm}_{\alpha}(\mathbf{k}),
\end{align}
along the same axes, as can be seen by taking the derivatives of the energy eigenvalues,
\begin{align} \label{eq:firstderivativeE}
\partial_{k_i} E^{\pm}_{\alpha}(\mathbf{k}) \sim & \pm\frac{\text{sgn}(k_i)\alpha|k_i|^{\alpha-1}}{2d(\mathbf{k}, \alpha)}.
\end{align}
The term \(|k_i|^{\alpha - 1}\) diverges along the momentum axes for \(0 < \alpha < 1\). In the following sections, we demonstrate that this regime leads to a divergent Berry curvature and quantum metric, indicating unphysical behavior. Despite this, it is essential to include this regime in our analysis, as it underscores the breakdown of our model and highlights the limitations of extending conventional Hamiltonians to fractional forms. For now, we proceed with our investigation of the Berry curvature and quantum metric. 

\subsection{Quantum geometric tensor}
A formulation of the  quantum geometric tensor  \cite{Liu2024} that avoids taking the derivatives of the Bloch states  \cite{Girvin}, $|\nabla_{\mathbf{k}}\psi_{n,\mathbf{k}}\rangle$, is given by

\begin{widetext}
\begin{align}
	\label{eq:QGT}
	Q_{ij}^n=\sum_{m\neq n}\frac{\langle \psi_{n,\mathbf{k}} |\partial_{k_i} H(\mathbf{k})| \psi_{m,\mathbf{k}} \rangle\langle \psi_{m,\mathbf{k}}|\partial_{k_j} H(\mathbf{k})|\psi_{n,\mathbf{k}}\rangle}{(E_n-E_m)^2},
\end{align}
\end{widetext}
where $n$ denotes the band index. The real and imaginary parts of Eq. \eqref{eq:QGT} define the quantum metric and the Berry curvature tensors, respectively:
\begin{align}
  g_{ij}^n =& \mathcal{R}e\{Q_{ij}^n\} \\
  \Omega_{ij}^n =& -2\mathcal{I}m\{ Q_{ij}^n\}.
\end{align}

For a two-band system \cite{Liu2024}, these quantum geometric components simplify to 
\begin{align}
\Omega^{\pm}_{ij} =& \mp\frac{1}{2d^3}(\partial_{k_i}\mathbf{d}\times\partial_{k_j}\mathbf{d})\cdot\mathbf{d} \\
g^{\pm}_{ij} =& \frac{1}{4d^2}[\partial_{k_i}\mathbf{d}\cdot\partial_{k_j}\mathbf{d}-\frac{1}{d^2}(\partial_{k_i}\mathbf{d}\cdot\mathbf{d})(\partial_{k_j}\mathbf{d}\cdot\mathbf{d})].
\end{align}

Owing to the simplicity of our model, we obtain the analytic result for the quantum geometric tensor of the fractional band dispersion:
\begin{align}
	\label{eq:fractionalQGT}
	Q_{\alpha,ij}^{\pm}(\mathbf{k}) = g_{\alpha,ij}^{\pm}(\mathbf{k}) - \frac{i}{2}\Omega_{\alpha,ij}^{\pm}(\mathbf{k}),
\end{align}
where the Berry curvature and quantum metric matrices are given by
\begin{widetext}
\begin{align}
	\label{eq:fractionalBerry}
	\mathbf{\Omega}_{\alpha}^{\pm}(\mathbf{k}) = \frac{m\alpha^2}{2d(\mathbf{k},\alpha)^3}
	\begin{pmatrix}
		0 & \mp |k_x k_y|^{\alpha-1} \\
		\pm|k_x k_y|^{\alpha-1} & 0
	\end{pmatrix}
\end{align}
and
\begin{align}
	\label{eq:fractionalMetric}
	\mathbf{g}_{\alpha}^{\pm}(\mathbf{k}) = \frac{\alpha^2}{4d(\mathbf{k},\alpha)^4}
	\begin{pmatrix}
		|k_x|^{2(\alpha-1)}(|k_y|^{2\alpha}+m^2) & -\text{sgn}(k_x)\text{sgn}(k_y)|k_x k_y|^{2(\alpha-1)} \\
		-\text{sgn}(k_x)\text{sgn}(k_y)|k_x k_y|^{2(\alpha-1)} & |k_y|^{2(\alpha-1)}(|k_x|^{2\alpha}+m^2)
	\end{pmatrix}.
\end{align}
\end{widetext}

For $\alpha=1$, $Q_{\alpha,ij}^{\pm}(\mathbf{k})$ reduces to the well-known quantum geometric tensor of the standard two-dimensional Dirac system \cite{Liu2024,Bernevig}.

\section{Berry curvature}\label{sec:berry}
We now examine the Berry curvature of the bands more closely. Since the Berry curvature of the valence band is simply the negative of that for the conduction band as can be seen in Eq. \eqref{eq:fractionalBerry}, it suffices to present the results for the conduction band only. In Fig. \ref{fig1:results} (e)-(h), we display the Berry curvature $\Omega_{\alpha,xy}^+$. The flat band (\(\alpha = 0\), Fig. \ref{fig1:results} (e)) shows a uniform Berry curvature of zero, which is consistent with the exact result Eq. \eqref{eq:fractionalBerry} showing that it is proportional to $\alpha^2$. The Dirac band (\(\alpha = 1\), Fig. \ref{fig1:results} (g)) exhibits the typical isotropic Berry curvature distribution. For \(0 < \alpha < 1\) (Fig. \ref{fig1:results} (f)), the Berry curvature is divergent along the \(k_x\)- and \(k_y\)-axes, where the corresponding energy bands exhibit non-analytic behavior.

This divergence can be understood by examining the analytic expression,
\begin{align}
	\label{eq:BerryAnalytic}
	\Omega_{\alpha,xy}^+ = -\frac{m\alpha^2|k_x k_y|^{\alpha-1}}{2(|k_x|^{2\alpha}+|k_y|^{2\alpha}+m^2)^{3/2}}.
\end{align}
Similar to the discussion following Eq. \eqref{eq:firstderivativeE}, the equation above contains $|k_i|^{\alpha-1}$ factors whose exponent is negative for $0<\alpha<1$. This characteristic gives the Berry curvature divergence along the momentum space axes in which $k_i=0$. Alternatively, the definition for the quantum geometric tensor in Eq. \eqref{eq:QGT} involves the derivatives of the fractional Bloch Hamiltonian,
\begin{align} \label{eq:firstderivativeH}
	\partial_i H_\alpha(\mathbf{k})\sim \text{sgn}(k_i) \alpha |k_i|^{\alpha - 1} \sigma^i \partial_{k_i} |k_i|,
\end{align}
which is also proportional to $|k_i|^{\alpha-1}$, a factor that diverges at $k_i=0$ for $0<\alpha<1$.

\subsection{Berry monopoles}
\begin{figure}[h!]
	\centering
	\includegraphics[width=0.7\linewidth]{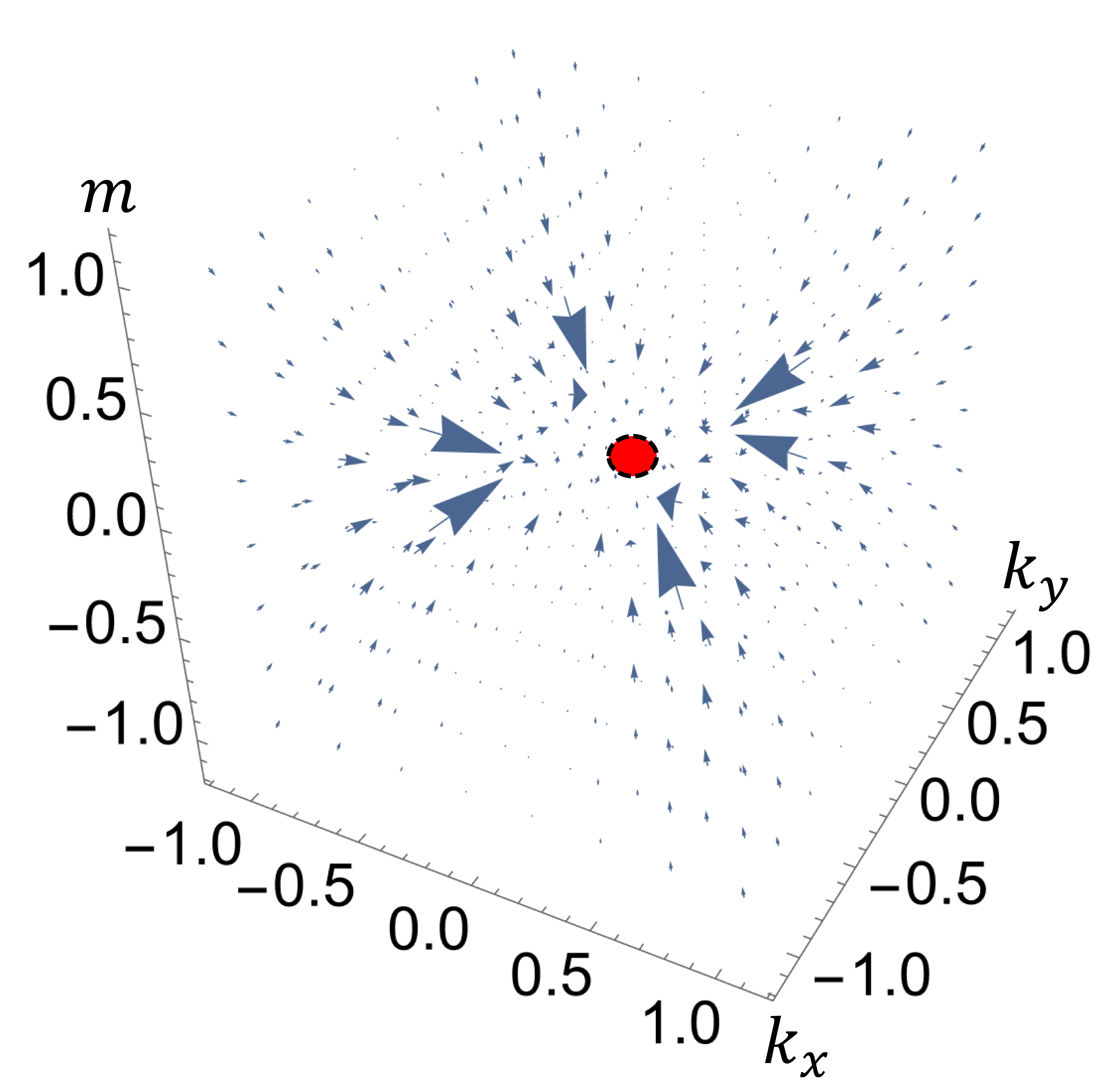}
	\caption{Vector plot of the pseudomagnetic field $\mathbf{B}_{\alpha}$ of the upper band in the $k_x-k_y-m$ parameter space for $\alpha=2.5$. The red dot is a guide to the eye and shows the sink of field lines.}
\label{fig2:monopole}
\end{figure}

We now analyze the formation of monopoles from the Berry curvature. First, we note that in the case of \(\alpha > 1\), the Berry curvature forms four \lq\lq petals\rq\rq located in each of the four quadrants of the \(k_x-k_y\) plane. As \(\alpha\) increases, these become more localized and concentrated at the location where the four sharp corners of the rectangular energy bands form in the momentum space. Their explicit locations are
\begin{align}
\label{berrypeak}
\pm k_x=\pm k_y=\left(\frac{\alpha-1}{2+\alpha}\right)^{\frac{1}{2\alpha}}m^{\frac{1}{\alpha}}.
\end{align}

In the Dirac limit $\alpha\rightarrow 1$, these four Berry curvature peaks converge towards the origin $k_x=k_y=0$, where the band touching occurs as the gap closes.

In conventional systems with integer dispersion, the formation of Berry curvature Dirac delta-like peak in the momentum space gives rise to the formation of the monopole. Let us therefore examine these four peaks that formed in the petals. Before we do, let us first briefly review the formation of Berry monopoles in a two-dimensional, two-band system following the References \cite{Xiao, Hasan2010, Fruchart2013IntroductionTI}. We can write the Hamiltonian as
    \begin{align}
    \label{2bandhamiltonian}
        H(\mathbf{k})=\Vec{d}(\mathbf{k})\cdot\Vec{\sigma}.
    \end{align}
In our analysis, we omit the term \( d_0(\mathbf{k})\mathbb{I}_{2\times 2} \), as it does not influence the eigenvectors and therefore has no impact on the band geometry. It is important to distinguish between two spaces involved in our discussion: the \(\mathbf{k}\)-space and the \(\vec{d}\)-space. Since we are dealing with two-dimensional systems, \(\mathbf{k} = (k_x, k_y)\) spans a two-dimensional manifold. In contrast, \(\vec{d}\)-space resides in three dimensions. The function \(\vec{d}(\mathbf{k})\) then defines a mapping from \(\mathbf{k}\)-space to \(\vec{d}\)-space, i.e., \(\mathbf{k} \mapsto \vec{d}\).
    
We now turn to the question of Berry monopoles, examining it from the perspectives of both \(\mathbf{k}\)-space and \(\vec{d}\)-space. In the two-dimensional \(\mathbf{k}\)-space, a Berry monopole manifests as a singularity in the Berry curvature—characterized by a Dirac delta-like peak—typically arising from a degeneracy or band crossing \cite{Fang2003AnomalousHE}. Alternatively, in the three-dimensional \(\vec{d}\)-space, the Berry monopole appears at the origin, \(\vec{d} = 0\). This correspondence becomes evident through the computation of the pseudomagnetic field, as we will detail for the fractional case below.

In systems with integer dispersion, these two perspectives are equivalent: in \(\mathbf{k}\)-space, Berry monopoles are located at band touching points where the Berry curvature exhibits a Dirac delta-like singularity; in \(\vec{d}\)-space, they correspond to the origin, \(\vec{d} = 0\).

While it is well established that in conventional systems the Berry curvature forms a sharp, Dirac delta-like peak at band touching points \cite{Fang2003AnomalousHE}, our analysis of the fractional case reveals a broader observation: Berry curvature tends to be concentrated at Brillouin zone locations of significant and non-analytic energy band surface curvature. In fact, from Eq. \eqref{berrypeak}, for $\alpha\rightarrow\infty$ and finite $m$, the Berry curvature becomes highly concentrated at the four isolated points $(k_x, k_y)=(\pm 1,\pm 1)$. This concentration of the Berry curvature to isolated points in $\mathbf{k}$-space is reminiscent of the formation of the Berry monopoles at the band crossings \cite{Fang2003AnomalousHE}. Do these highly concentrated Berry curvature peaks represent true monopoles in the sense that they carry topological charge in $\vec{d}$ space? We will demonstrate below that they do not, and that, surprisingly, the topological charge remains located at the origin of $\vec{d}=0$, which corresponds to the point $\mathbf{k}=0$ in the Brillouin zone.

For the two-band Hamiltonian Eq. \eqref{2bandhamiltonian}, the Hodge dual of the Berry curvature $B_i\equiv\varepsilon_{ijk}\Omega^{jk}/2$ gives 
\begin{align}
\mathbf{B}=\frac{1}{2}\frac{\hat{\mathbf{d}}}{d^2},
\end{align}
which is the pseudo-magnetic field produced by a monopole at the origin $\vec{d}=0$.

For our model, the x and y components of $\vec{d}$ are given by Eqs. \eqref{d1} and \eqref{d2}, respectively. The z component is then $d_3=m$. Following the calculations above, the corresponding pseudo-magnetic field for our model is
\begin{align}
\mathbf{B}_{\alpha}=&-\frac{\text{sgn}(k_x)\alpha|k_x|^{\alpha } |k_y|^{\alpha -1}}{\left(|k_x|^{2 \alpha }+|k_y|^{2 \alpha }+m^2\right)^{3/2}}\hat{\mathbf{k}}_x \nonumber\\
&-\frac{\text{sgn}(k_y)\alpha|k_x|^{\alpha -1} |k_y|^{\alpha }}{\left(|k_x|^{2 \alpha }+|k_y|^{2 \alpha }+m^2\right)^{3/2}}\hat{\mathbf{k}}_y\nonumber\\
&-\frac{m\alpha ^2  |k_xk_y|^{\alpha -1}}{\left(|k_x|^{2 \alpha }+|k_y|^{2 \alpha }+m^2\right)^{3/2}}\hat{\mathbf{k}}_z.
\end{align}
Although the magnitude of the magnetic field is greatest at the points described by Eq. \eqref{berrypeak}, the monopole at $\vec{d}=0$, which serves as the source or sink of \(\mathbf{B}_{\alpha}\), is still located at the origin $\mathbf{k}=0$ as shown in Fig. \ref{fig2:monopole}.

\subsection{Chern number}
\begin{figure}[h!]
	\centering
	\includegraphics[width=1\linewidth]{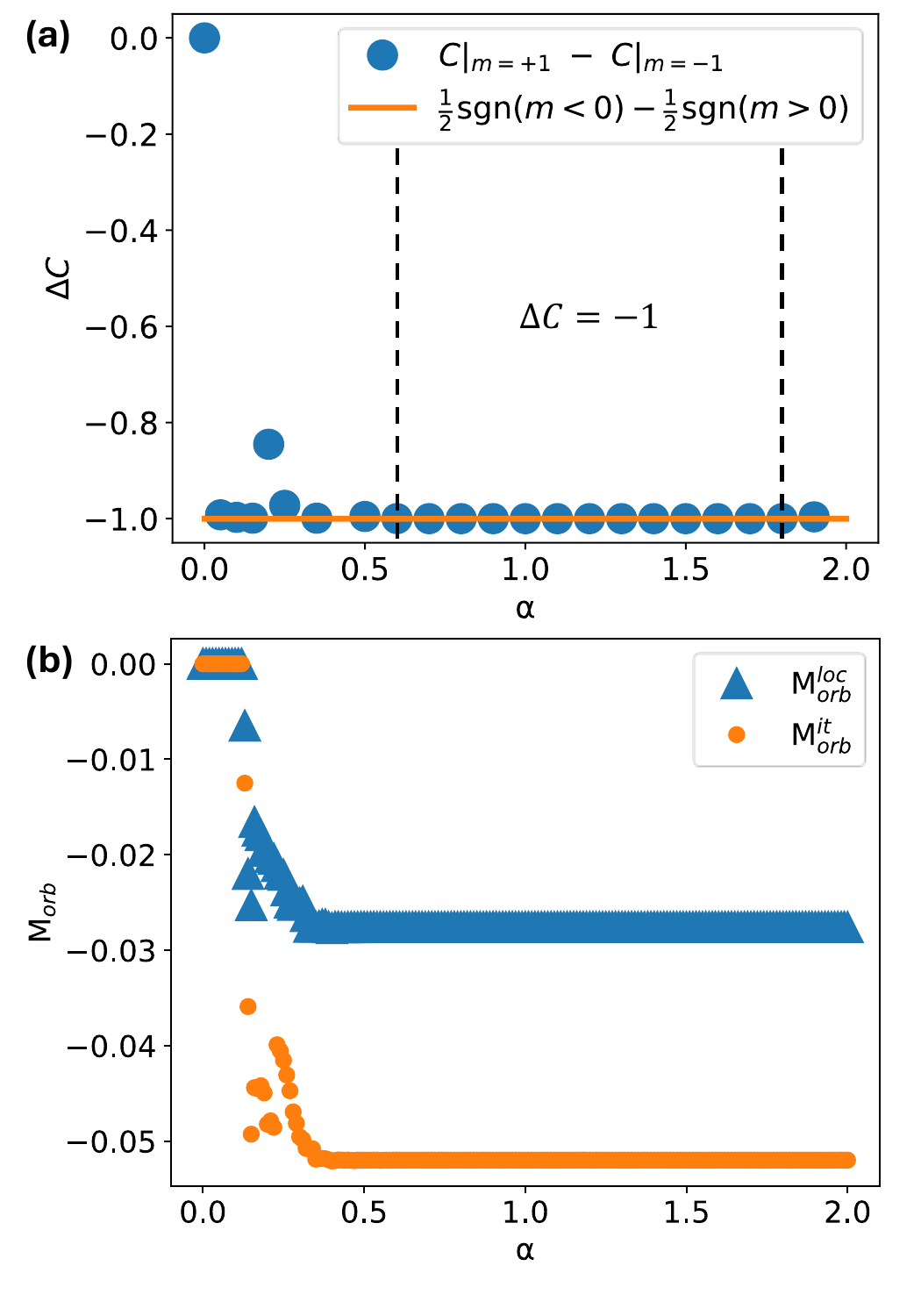}
	\caption{(a) Difference in $C$ between $m=+1$ and $m=-1$, over $\alpha \in [0, 2]$. The solid orange line is the change in the Hall conductance across $+m$ to $-m$ for a Dirac fermion in the continuum. The region within the dashed lines denotes the range of $\alpha$ where $\Delta C$ was evaluated to be exactly $-1$. (b) Plot of the local $M_{orb}^{loc}$ and the itinerant contributions $M_{orb}^{it}$ to the orbital magnetization for $\mu=1$ over $\alpha \in [0, 2]$.}
	\label{fig3:BerryIntegral}
\end{figure}

We now investigate the integral of the Berry curvature over momentum space:
\begin{align} \label{eq:Hall}
C = \frac{1}{2\pi} \int_{-\infty}^{\infty} \int_{-\infty}^{\infty}dk_x dk_y\Omega_{xy}(\mathbf{k}, m, \alpha).
\end{align}
This integration is performed over the entire \(\mathbf{k}\)-space since our model, as described by Eq. \eqref{hamiltonian}, is based on the continuum Dirac Hamiltonian \cite{Bernevig, Vanderbilt}. As a result, the quantity defined in Eq. \eqref{eq:Hall} is not quantized to integer values and technically does not represent a Chern number, which is only defined for compact \(\mathbf{k}\) spaces. However, the \textit{change} in \(C\) across a possible topological transition, as the gap varies from \(m = -1\) to \(m = 1\), can still be interpreted as a change in the Chern number across this transition \cite{Bernevig, Vanderbilt}. 

We compute this change as follows:
\begin{align} 
\label{eq:deltaHall}
\Delta C = C|_{m=+1} - C|_{m=-1}.
\end{align}
As shown in Fig. \ref{fig3:BerryIntegral} (a), we find that \(\Delta C\) is quantized to the integer value \(-1\), even for non-integer values of \(\alpha > 0.5\). In the interval \(0 < \alpha < 0.5\), however, the Berry curvature spreads along the axes, requiring a larger momentum region for integration to accurately capture the change in the Chern number. This explains the observed deviation from \(\Delta C = -1\) in the figure. We investigate this change in Chern number further by deriving the closed-form formula for $C$.

For $\alpha=0$, Eq. \eqref{eq:Hall} trivially vanishes since the integrand is zero, as can be seen from Eq. \eqref{eq:fractionalBerry}. We therefore focus on the case $\alpha \neq0$. We obtain a closed-form expression by taking advantage of the C4-symmetry of the Berry curvature as established by Fig. \ref{fig1:results}. We get the total integral of $\Omega_{\alpha, xy}^+$ by quadrupling the contribution of the first quadrant. In this region, Eq. \eqref{eq:BerryAnalytic} simplifies to 
\begin{align}
    \Omega_{\alpha, xy}^+=-\frac{m\alpha^2(k_x k_y)^{\alpha-1}}{2(k_x^{2\alpha}+k_y^{2\alpha}+m^2)^{3/2}}.
\end{align}
The expression for $C$ becomes
\begin{align}
\label{eq:HallSimplified}
	C&=-\frac{2}{\pi} \int_{0}^{\infty} \int_{0}^{\infty} dk_xdk_y \frac{m\alpha^2(k_x k_y)^{\alpha-1}}{2(k_x^{2\alpha}+k_y^{2\alpha}+m^2)^{3/2}}\\
\label{eq:HallEvaluation}
	&=-\frac{2}{\pi} \int_{0}^{\infty} dk_y \frac{m\alpha k_y^{\alpha-1}}{2(k_y^{2 \alpha}+m^2)}\\
	&=-\frac{1}{\pi} \tan ^{-1}\Bigg(\dfrac{k_y^{\alpha}}{m}\Bigg) \Bigg|_0^{\infty}\\
	&=-\frac{\text{sgn}(m)}{2},
\end{align}
which shows that $\alpha$ drops out in the final expression. This gives a quantized change in Chern number Eq. \eqref{eq:deltaHall}
\begin{align}
    \Delta C=-1,
\end{align}
independent of $\alpha$, provided that $\alpha>0$. Note that this result is similar to that of the standard Dirac Hamiltonian $\alpha=1$ \cite{Bernevig}, thereby showing that the fractional dispersion does not change the topological property $\Delta C$.

For \(\alpha = 0\), the situation is different. As seen in Fig. \ref{fig1:results} (e), the Berry curvature is uniformly zero across momentum space, remaining so even as \(m\) varies from negative to positive. This can be seen in general from Eq. \eqref{eq:fractionalBerry}, where the Berry curvature is shown to be proportional to $\alpha^2$. This results in \(\Delta C = 0\), indicating that there is no topological transition when \(m\) is varied across zero for \(\alpha = 0\). This is consistent with our previous discussion, where we noted the absence of band gap closing at \(m = 0\) for \(\alpha = 0\).

\subsection{Orbital magnetization}\label{sec:orbital}
\begin{figure*}[htb!]
	\centering
	\includegraphics[width=1\linewidth]{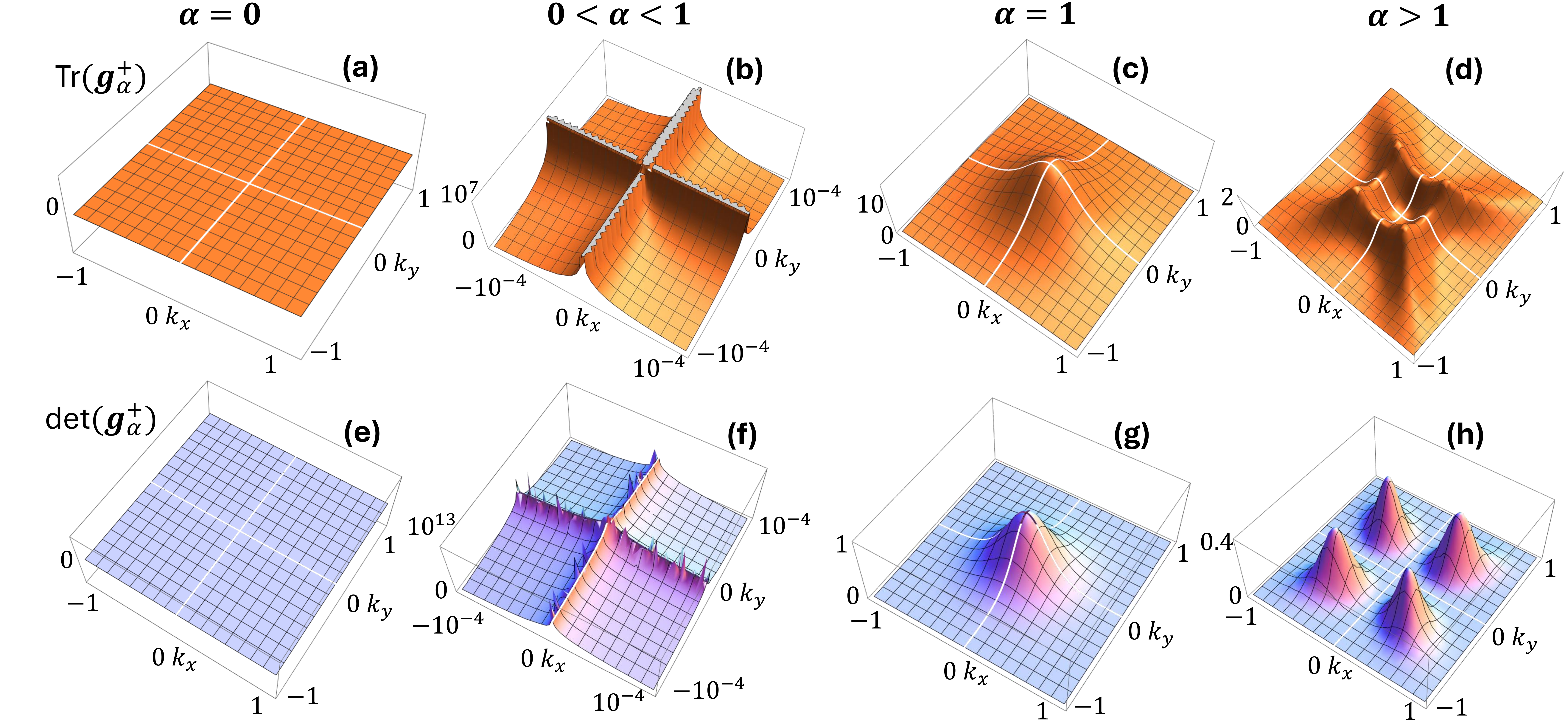}
	\caption{The plots illustrate the trace (a)-(d) and the determinant (e)-(h) of the quantum metric for the conduction band across different values of \(\alpha\), with the band gap parameter set to \(m=0.5\). In panels (b) and (f), we used \(\alpha=0.3\), with the domain of the plots zoomed in and the range displayed on a logarithmic scale to clearly highlight the divergence along the momentum space axes. In panels (d) and (h), we used \(\alpha=2.5\).}
	\label{fig4:results}
\end{figure*}

To explore the physical consequence of the Berry curvature redistribution, we calculate the orbital magnetization \cite{Ceresoli2006, Thonhauser2005, Thonhauser2011}, a quantity that is accessible in experiments. In magic-angle graphene, for example, the orbital magnetism can be imaged using a scanning superconducting quantum interference device \cite{Grover2022}. We have two contributions: the local orbital magnetization
\begin{align}
    \mathbf{M}_{orb}^{loc}=\frac{1}{(2\pi)^2}\mathcal{I}m\sum_n\int_{E_{n\mathbf{k}}\leq\mu}d^2k\langle\partial_\mathbf{k}\psi_{n,\mathbf{k}}|\times H_\mathbf{k}|\partial_\mathbf{k}\psi_{n,\mathbf{k}}\rangle
\end{align}
and the itinerant contribution
\begin{align}
    \mathbf{M}_{orb}^{it}=&\frac{1}{(2\pi)^2}\mathcal{I}m\sum_n\int_{E_{n\mathbf{k}}\leq\mu}d^2k\langle\partial_\mathbf{k}\psi_{n,\mathbf{k}}|\times(E_{n,\mathbf{k}}\nonumber\\
    &-2\mu)|\partial_\mathbf{k}\psi_{n,\mathbf{k}}\rangle,
\end{align}
where $\mu$ is the chemical potential. Here, we consider the conduction band only. In conduction bands, the total orbital magnetization $M_{orb}$ was shown to be proportional to the integral of the Berry curvature \cite{Xiao}.

We show our results on Fig. \ref{fig3:BerryIntegral} (b). The orbital magnetization is constant in the range $\alpha\gtrsim 0.4$. This is because while the Berry curvature is redistributed inside the momentum region $E_\mathbf{k}\leq\mu$ as $\alpha$ is varied, it does not spread outside the integration region, rendering the integral constant. We found that this constant magnetization persists for larger $\alpha$, but the numerical integration must be refined as the Berry curvature becomes highly localized into four monopoles. For $0<\alpha <0.4$, some of the Berry curvature leaks out of the integration region as shown in Fig. \ref{fig1:results} (f), resulting in decreased orbital magnetization magnitude. At exactly $\alpha=0$, the orbital magnetization contributions are zero since the Berry curvature vanishes identically in the entire momentum space as previously shown in Fig. \ref{fig1:results} (e).

It is interesting to compare our model and magnetization results with those of chirally stacked graphene \(N\)-layer systems, which exhibit a large exponent \(N\) dispersion. Furthermore, as mentioned in the introduction, the application of an external electric field perpendicular to the layers can interpolate the dispersion between two integer values \cite{Dong2023}, potentially enabling the realization of fractional dispersion in multi-layer graphene systems.

In contrast to our finding that the magnetization is independent of \( \alpha \), the magnetization in chirally stacked graphene is proportional to the exponent \( N \) \cite{Zhang2011}. This discrepancy originates from the difference in the underlying Bloch Hamiltonians. The Hamiltonian for chirally stacked graphene is given by
\begin{align}
\label{stackedgraphene}
    H_N=\frac{(v_0p)^N}{(-\gamma_1)^{N-1}}[\cos(N\phi_\mathbf{p})\sigma_x+\sin(N\phi_\mathbf{p})\sigma_y]+\lambda\sigma_z,
\end{align}
where $\cos\phi_\mathbf{p}=\tau_xp_x/p$ and $\sin\phi_\mathbf{p}=p_y/p$.

In contrasts to our simpler 2 $\times$ 2 Hamiltonian model, Eq. \eqref{stackedgraphene} includes the layer pseudospin $\vec{\sigma}$, valley $\vec{\tau}$, as well as real spin $\vec{s}$, which gives rise to three distinct symmetry-broken ground states by replacing the term $\lambda\sigma_z$ with $\lambda\tau_z\sigma_z$, $\lambda s_z\sigma_z$, or $\lambda\tau_zs_z\sigma_z$.

For $N=1$, the model \eqref{stackedgraphene} reduces to two copies of Dirac Hamiltonian which is similar to our model when $\alpha=1$. The difference becomes more pronounced for $N=2$, where Eq. \eqref{stackedgraphene} becomes
\begin{align}
    H=\frac{v_0^2}{(-\gamma_1)}[(p_x^2-p_y^2)\sigma_x+2p_xp_y\tau_z\otimes \sigma_y]+\lambda\sigma_z.
\end{align}

It would be interesting to study the fractional generalization of the stacked graphene model:
\begin{align}
    H_\alpha=\frac{(v_0p)^\alpha}{(-\gamma_1)^{\alpha-1}}[\cos(\alpha\phi_\mathbf{p})\sigma_x+\sin(\alpha\phi_\mathbf{p})\sigma_y]+\lambda\sigma_z,
\end{align}
where $\alpha$ is not restricted to integer values, which we leave for future study.

\section{Quantum metric}\label{sec:metric}
We now examine the results of the quantum metric. This gives us the information about the overlap of the states in $\mathbf{k}$-space. We show on Fig. \ref{fig4:results} the determinant and trace of the quantum metric which provide avenues for probing quantum geometry in experiment. The determinant supports an alternative method for characterizing topological invariants \cite{Tan2019}, while the trace has been demonstrated to measure the band drude weight through directional lattice oscillations \cite{Kang2024, Ozawa2018}. It is sufficient to show the metric invariants for the upper band since the lower band have identical quantum metric components as evident in Eq. \eqref{eq:fractionalMetric}. For the flat band, $\alpha=0$, the quantum metric is zero everywhere in the momentum space as shown in Fig. \ref{fig4:results} (a) and (e). This tells us that all the states have complete overlap and are therefore identical. We found similar results when $m$ is varied across $m=0$, which is consistent with previous findings that there is no band gap closing, Fig. \ref{fig1:results} (a), and that there is no change in the Chern number, Fig. \ref{fig3:BerryIntegral} (b). Hence, for the flat band case, our model describes a trivial insulator for all values of $m$.

For non-integer dispersions, $0<\alpha<1$, the quantum metric components are divergent along the momentum space axes as shown on Fig. \ref{fig4:results} (b) and (f). This mirrors the non-analytic regions of the energy band dispersion and the Berry curvature, Figs. \ref{fig1:results} (b) and (f), respectively. This can be understood by examining the exact forms for the trace and the determinant of the quantum metric,
\begin{align}
	\label{eq:trace}
	\text{Tr}(\mathbf{g}_{\alpha}^{\pm})=&\frac{\alpha^2}{4 d(\mathbf{k},\alpha)^4} \bigg[|k_x|^{2(\alpha-1)}(|k_y|^{2\alpha}+m^2) \nonumber\\
	&+ |k_y|^{2(\alpha-1)}(|k_x|^{2\alpha}+m^2)\bigg]
\end{align}
\begin{align}
	\label{eq:determinant}
	\text{det}(\mathbf{g}_{\alpha}^{\pm})=&\frac{m^2\alpha^4}{16 d(\mathbf{k},\alpha)^8} \bigg[m^2 |k_x k_y|^{2(\alpha-1)}\nonumber\\
	&+|kx|^{2(2\alpha-1)}|ky|^{2(\alpha-1)}\nonumber\\ &+|ky|^{2(2\alpha-1)}|kx|^{2(\alpha-1)}\bigg].
\end{align}
These expressions exhibit terms that are proportional to $|k_i|^{\alpha-1}$ which diverge along the momentum space axes $0<\alpha<1$. Just as in the case of the Berry curvature, the divergence of the quantum metric is also attributed to the divergent first derivatives over the axes of the fractional Bloch Hamiltonian for $0<\alpha<1$.

For $\alpha>1$, Figs. \ref{fig4:results} (d) and (h), we found four smooth finite peaks, one for each quadrant in the momentum space. These peaks are found at the same locations of the four \lq\lq petals\rq\rq of the Berry curvature in Fig. \ref{fig1:results} (h). The corresponding locations between the peaks of the quantum metric and the petals of the Berry curvature may be ascribed to the equalities,
\begin{align} \label{eq:tr}
    \text{Tr}(\mathbf{g^{\pm}})&=|\Omega_{xy}^{\pm}| \\ \label{eq:det}
    \sqrt{\text{det}(\mathbf{g^{\pm}})}&=\frac{|\Omega_{xy}^{\pm}|}{2},
\end{align}
that are unique for two-band systems \cite{Ozawa2021, Liu2024}. Consistent with Eqs. \eqref{eq:tr} and \eqref{eq:det}, we found for all cases of $\alpha$ that the peaks in the quantum metric, just like that of the Berry curvature, occur at the locations where the energy band curvature is high.

For the massive Dirac band case shown in Figs. \ref{fig4:results} (c) and (g), $\alpha=1$, we found the typical isotropic and smooth peak at the center of the momentum space. 

While the transition from a finite metric for \(\alpha > 1\) to a divergent metric for \(\alpha < 1\) is found in the topological phase transitions in models like the Su-Schrieffer-Heeger model \cite{Chen2024}, we will argue that in our model, the regime \(0 < \alpha < 1\) may be unphysical.

\section{Conclusions}\label{sec:conclusion}
In summary, our calculations reveal that non-integer dispersion introduces significant and novel effects on quantum band geometry. Specifically, we observe that fractional dispersion redistributes both the Berry curvature and the quantum metric. These quantities become highly concentrated in momentum space regions where the energy band exhibits strong curvature, such as at the corners, for \(\alpha > 1\).

We now give our final comments for the region \( 0 < \alpha < 1 \), reflecting from all the results that we have obtained. In this range, we observe that both the Berry curvature and the quantum metric diverge along the momentum axes. Additionally, the Fermi velocity, as described by Eqs. \eqref{eq:velocity} and \eqref{eq:firstderivativeE}, as well as the momentum gradient of the Bloch eigenfunctions, also exhibit both divergences and discontinuities. The divergence in the latter implies that the center positions of the Wannier functions is located at infinity. These findings strongly indicate that the regime \( 0 < \alpha < 1 \) may be unphysical. This conclusion is consistent with the path integral formulation of fractional quantum mechanics, where Brownian trajectories are replaced by L\'evy flights. In this framework, the region \(0 < \alpha < 1\) is excluded, as it leads to a divergent first moment \cite{Stickler2013}. Notably, even without invoking this framework of fractional quantum mechanics, the unphysical nature of this regime manifests as a discontinuity in the Fermi velocity and a divergence in the quantum geometric tensor.

It is well-established that Berry monopoles form at band touching points, with two key signatures. The first, and most significant, is the formation of topological charge in $\vec{d}$ space, which acts as a source or sink for the pseudo-magnetic field. The second is that the Berry curvature vanishes almost everywhere in the band, becoming concentrated at the band touching points. Our results reveal a separation of these two effects for fractional systems (\(\alpha > 1\)). While the Berry monopole, corresponding to the topological charge at $\vec{d}=0$, remains located at the origin $\mathbf{k}=0$, the Berry curvature develops Dirac delta-like peaks at four distinct points, away from $\mathbf{k}=0$.

It is truly remarkable that quantum systems with fractional dispersion can be realized in a variety of media. In this work, we have demonstrated that the interplay between fractional quantum mechanics and quantum geometry leads to rich phenomena not observed in conventional systems with integer dispersion. The quantum geometric tensor, particularly its quantum metric component, remains largely unexplored even in traditional systems. This presents an exciting opportunity to investigate the physical implications of fractional dispersion on geometry. Future directions of research could include exploring the effects on Hall transport, superfluid weight in multi-band superconductors, and non-linear optical responses, among many other phenomena. We hope this work will inspire further investigations into the connections between fractional systems and geometry.

\bibliography{apssamp}

\end{document}